\documentclass{svjour3}                     
\smartqed  
\usepackage{graphicx}

\usepackage{amsfonts}
\usepackage{amssymb}
\usepackage{latexsym}
\usepackage{color}
\usepackage{layout}
\usepackage[english]{babel}
\usepackage[nointlimits]{amsmath}
\usepackage{bbding}
\newcommand{\mc}[1]{{\mathcal #1}}

\newcommand{\bb}[1]{{\mathbb #1}}

\newcommand{\ol}[1]{\,\overline {\!#1\!}\,}
\newcommand{\ul}[1]{\underline {#1}}

\renewcommand{\epsilon}{\varepsilon}
\newcommand{\id}{{1 \mskip -5mu {\rm I}}}

\newcommand{\p}{\par}
\newcommand{\n}{\noindent}

\newcommand{\virgolette}[1]{``#1''}

\newcommand{\dint}{\displaystyle \int}

\renewcommand{\epsilon}{\varepsilon}
\newcommand{\capa}{\mathop{\textrm{cap}}\nolimits}

\def\JELname{\textbf{JEL Classification}\enspace}
\def\JEL#1{\par\addvspace\medskipamount{\rightskip=0pt plus1cm
\def\and{\ifhmode\unskip\nobreak\fi\ $\cdot$}\noindent\JELname\ignorespaces#1\par}}

\begin{document}

\title{Modelling interest rates by correlated multi-factor CIR-like processes
}


\author{Lorenzo Bertini        \and
        Luca Passalacqua 
}


\institute{Lorenzo Bertini \at
              Dipartimento di Matematica,
              Universit\`a di Roma ``La Sapienza''\\
              Piazzale A. Moro 2, 00185 Roma (Italy) \\
              \email{bertini@mat.uniroma1.it}           
           \and
              Luca Passalacqua (\Envelope) \at
              Dipartimento di Scienze Attuariali e Finanziarie,
              Universit\`a di Roma ``La Sapienza'', \\
              Via Nomentana 41, 00161 Roma (Italy) \\
              Tel.: +39-06-49919559\\
              Fax: +39-06-44250289\\
              \email{luca.passalacqua@uniRoma1.it}           
}


\maketitle

\begin{abstract}
We investigate the joint description of the interest-rate term stuctures
of Italy and an AAA-rated European country by mean of a --here proposed--
correlated  CIR-like bivariate model where one
of the state variables is interpreted as a benchmark risk-free rate and
the other  as a credit spread.
The model is constructed by requiring the strict positivity of interest
rates and the asymptotic decoupling of the joint distribution of the
two state variables on a long time horizon. The second condition is met by imposing
the reversibility of the process with respect to a product measure, the first is
then implemented by using the tools of potential theory.
It turns out that these conditions select
a class of non-affine models, out of which we choose one that
is quadratic in the two state variables both in the drift and
diffusion matrix. We perform a numerical analysis of the model
by investigating a cross section
of the term structures comparing the results with those obtained with an uncoupled bivariate CIR model.


\JEL{E43}
\end{abstract}

\section{Introduction}
\label{intro}
The difficulty to model the evolution of the term structure of
interest rates is witnessed by the existence of a large number
of models present in the academic literature and in the financial
practice, see \emph{e.g.} \cite{Brigo,ShreveII} for a review. Broadly speaking, these models can be grouped in
financially oriented \emph{arbitrage} models, whose main objective is pricing
interest rate sensitive contracts and measuring risk associated
with the time evolution of the term structure, and economically oriented
models that are embedded in more complex market equilibrium models.
Among equilibrium models that of Cox, Ingersoll and Ross (hereafter CIR)
is certainly one of the most attractive. This model, introduced in \cite{CIRa,CIRb},
is characterized by two main properties:
mean-reversion to an asymptotic state and absence of
negative interest rates.
Moreover, as Gaussian-like models (\emph{i.e.} models founded on Ornstein-Uhlenbeck
processes) generally develop numerically relevant tails in region of
negative interest rates with growing time horizons, the CIR formulation is
particularly popular in financial applications having as underlying
portfolios composed of government bonds and long time horizons, such as
the strategic asset allocation of life insurance segregated funds.
However, well known limits of the CIR model are that the term structure can
assume (see, \emph{e.g.} \cite{Kan}) only the following three shapes:
monotonically increasing,
monotonically decreasing and humped ({\it i.e.} increasing to a maximum
and then decreasing), the need to allow the model parameters to vary with
time in order to capture the observed evolution (see, \emph{e.g.} \cite{Brigo}),
and the difficulty to  describe simultaneously all types of interest rate
sensitive contracts, such as interest rate swaps, caps and swaptions
(see, \emph{e.g.} \cite{scs}).
Moreover a single factor model is unable to describe simultaneously
the evolution of the term structure of real and nominal interest rates.

All the above difficulties lead quite naturally to multi-factor extensions
of the basic univariate CIR model.
For example, already in the original model proposed by Cox, Ingersoll
and Ross in \cite{CIRb}, the instantaneous
nominal interest rate is a linear combination of two
independent state variables, the real interest rate and the
expected instantaneous inflation rate, each evolving in time
according to univariate diffusion processes, thus realizing the
stochastic version of the well-known Fisher equation.
Another example is the two-factor extension proposed by Longstaff and
Schwartz \cite{LS}, where the two
factors are used to express the short rate and its
volatility. A different interpretation proposed for the two factor model
is that the factors are linked to the short and long
(w.r.t. the maturity of the contract) rates, as
in the Brennan and Schwartz model \cite{BS}. A three-factor extension
has also been considered and empirically investigated, among others,
by Chen and Scott \cite{ChenScott} on U.S. market data. The three factor
setting is often motivated by the findings of Litterman and Scheinkman \cite{LiS}
according to whom the empirical description of the
intertemporal variation of the term structure needs the use of three
factors: the general level of interest rates, the slope of the yield curve
and its curvature, that is associated with the volatility.
For the euro market, a recent empirical investigation of the term structure evolution
\cite{Passalacqua} has shown that two factors are sufficient for
a description of the data with mean squared error Gaussianly distributed
with about 10 basis points dispersion around the observed values.

The aim of this paper to investigate the interest-rate spread
between the Government debt
of two selected European Union member states, Germany and Italy,
in the hypothesis that the spread reflects the different market opinions
of their respective credit quality.
Both countries, together with France, are known to possess the most liquid
and high-volume Government bond markets in Europe which provide observations
for a broad maturity range so that it seems reasonable to assume that
the impact of liquidity premia in bond prices can be safely neglected.
To model the joint term structures of interest rates
we introduce a two-factor CIR-like model where one
of the factors is interpreted as a benchmark risk-free rate and
the other is a credit spread.
In this sense the model follows the fractional recovery approach of
Duffie and Singleton \cite[\S 7.2]{Duffie},
although --as discussed later-- our model is not affine in the
state variables.
In fact, since it is natural to expect that the same macroeconomic factors
affects both the level of interest rates and credit spreads, it
is unclear to what extent a two independent factor model could
describe the joint behaviour of the Italian and German rates.
This issue is particularly relevant in the measurement of
risk measures on portfolios composed by Italian
and AAA-rated (\emph{e.g.} German) government bonds.

The two-factor model investigated in this paper is costructed according to
the following requirements.

First of all, from the financial point of view, a fundamental requirement of
nominal interest rates modelling is
to avoid negative interest rates. In the univariate CIR model this is
guaranteed by the choice of the stochastic differential equation.
However, since interest rates are expected to be strictly positive, it is also important
to establish under which conditions on the parameters the rates do not vanish.
In the case of the single factor CIR model, this question has been
solved, in a different context, by Feller \cite{Feller}, obtaining
a necessary and sufficient condition.
More generally, the hitting conditions for one-dimensional diffusion
processes have been completely characterized , see \emph{e.g.} \cite{RevuzYor}.
On the other hand, the multi-dimensional case is much less understood.
For financially oriented multi-factor models this question has been
partially addressed in \cite{DuffieKan}.
From a mathematical point of view, potential theory methods, when
applicable, are the natural tools to analyze the hitting conditions
for diffusion processes \cite{Fukushima}. In fact, they have extensively been used
in several contexts, albeit -- in our knowledge --
not for financial applications.
In this paper we analyze by these methods the hitting conditions for multivariate
correlated CIR-like processes and apply the result in the costruction
of the correlated two-factor model.

The second important feature we require on the model is that the correlation asymptotically
vanishes. More precisely, we impose that in the limit of infinitely far time horizon the joint
distribution of the benchmark risk-free rate and the credit spread decouples into the product of two Gamma distributions.

Finally, among the models meeting the above requirements, we select the \virgolette{minimal} class,
by further requiring that the drift and diffusion matrix are quadratic in the state variables.
We shall refer to this model as the \virgolette{asymptotically decoupling correlated}
model or ADC model.

As previously stated, we investigate the ability of the ADC model to capture market behavior
by applying the model to the joint description
of Italian and German term structures of interest rates, at a fixed calendar date.
In the application the numerically demanding calibration of the ADC model, for which there is
no closed form expression for discount factors, has been
performed with the well known technique of simulated annealing,
in the (fast) adaptive version developed by Ingber \cite{Ingber}.
Finally, we compare the results obtained with the ADC model with those obtained with
a \virgolette{simple} bivariate CIR model with uncoupled state variables.

Our finding is that, for the particular cross section here examined, the ADC model does not increase significantly
the accuracy in the description of the two term structures with respect to
the one achieved by the uncoupled bivariate CIR model.
However, the predicted risk-neutral joint distribution of the two models are
different. This suggests that --once risk premia are inferred from
the analysis of time series--
the \virgolette{natural} distributions could be different, implying
different values of risk measures for the same portfolios.

\section{The univariate CIR process}
\label{sec:2}

In this section we introduce the tools of potential theory by discussing in a
self-contained way the condition for the strict positivity of the univariate CIR process.
Fix a filtered probability space $\big( \Omega, \mc F, \mc F_t,\mathbb P \big)$
equipped with a standard Wiener process $w$.
The univariate CIR process $X=\{X_t\,\: t\in \bb R_+\}$ is defined as the solution to the Ito equation
\begin{equation}
\label{CIR1}
\begin{array}{lcl}
d X_t &=&  \kappa [\theta - X_t] dt + \sigma \, {\displaystyle
  \sqrt{X_t}} \, dw_t \\
X_0 &=& x_0
\end{array}
\end{equation}
where $\kappa,\theta,\sigma$ are strictly positive parameters and $x_0 > 0$ is
the initial condition.
In the celebrated paper of Feller \cite{Feller} it is shown that the transition
probability density of $X_t$ is given by
\begin{equation}
p_t(x_0,x) =
c \, e^{-(u+v)} \, \biggl( \sqrt{\dfrac{v}{u}} \biggr)^{\nu-1} \, I_{\nu-1}(2\sqrt{uv})
\label{CIRpdf}
\end{equation}
where
\begin{equation}
\begin{array}{cccc}
c = \dfrac{2\kappa}{\sigma^2 (1-e^{-\kappa t})}, & u = c \, x_0 \,e^{-\kappa t},
& v = c \, x, & \nu = \dfrac{2 \kappa \theta}{\sigma^2}.
\end{array}
\end{equation}
and $I_{\alpha}$ is the modified Bessel function of the first kind of
order $\alpha$.
Notice that at fixed time $t$ the dependence of $p_t$ on $x$ is only
through $v$, while $c$, $u$ and $\nu$ are constant parameters.
The parameter $\nu$ controls the behaviour of the probability density in \eqref{CIRpdf}
as $x \! \downarrow \! 0$: for $\nu > 1$ $p_t(x_0,x)$
vanishes, for $\nu=1$ it converges to $c e^{-u}$, while for $0< \nu < 1$ it
behaves as $x^{\nu-1}$ and therefore it has a integrable singularity.
In fact, the properties of the modified Bessel function for $\alpha>-1$
are such that $I_{\alpha}(y)$ is real and positive for any $y>0$ and
that in the limit $y\downarrow 0$ one has, see e.g.\
\cite[pp.~374]{Abramowitz},
\begin{equation}
I_{\alpha}(y) = \frac{1}{\Gamma(\alpha+1)}\Bigr(\frac{y}{2}
\Bigl)^{\alpha} +
\frac{1}{\Gamma(\alpha+2)}\Bigr(\frac{y}{2} \Bigl)^{\alpha+2} + O(y^4)
\label{AS9.6.7}
\end{equation}
where $\Gamma(z)=\int_0^{\infty} t^{z-1} e^{-t}dt$ is the Euler gamma function.
The cumulative distribution function $F(x,t) = {\mathbb P}_{x_0}[X_t
\le x]$ is obtained by integrating \eqref{CIRpdf}
\begin{equation}
F_t(x_0,x) = \int_{0}^{x}\!dy\, p_t(x_0,y)
 = \int_{0}^{c \, x}\! dy \, e^{-(u+y)} \biggl( \sqrt{\dfrac{y}{u}} \biggr)^{\nu-1}
I_{\nu-1}(2\sqrt{uy}) =  \tilde{\chi}^2(2cx;2\nu,2u)
\end{equation}
where $\tilde{\chi}^2(x;n,\lambda)$ is the cumulative
distribution function of a non-central chi-square distribution
with $n$ degrees of freedom and non-centrality parameter $\lambda$.

The long time behaviour of the real random variable $X_t$ is given by
the probability density $\pi_\nu(x) = \lim_{t\to\infty}
p_t(x_0,x)$. It is simple to check that this limit is independent on
the initial condition $x_0$ and $\pi_\nu$ is just the density of a
Gamma distribution with parameters $\nu=2 \kappa \theta/\sigma^2$ and ${\omega}=\nu/\theta$, namely
\begin{equation}
\label{misinv}
\begin{array}{l}
\pi_\nu (x) =  N \, x^{\nu -1} e^{-\omega x} =N \, x^{\nu -1} e^{- \nu x /\theta} \\
N := \dfrac{\omega^{\nu}}{\Gamma(\nu)} = \Big(\frac{\nu}{\theta}\Big)^{\nu} \frac{1}{\Gamma(\nu)}
\end{array}
\end{equation}

We give now a potential theoretical proof of the classical result, again due to Feller
\cite{Feller}, that the CIR process hits the origin iff $\nu\ge 1$.
Referring to \cite{Fukushima} for an exhaustive treatment, we recall
the basic notions of potential theory of reversible Markov
process.

The generator of the process $X$, solution to \eqref{CIR1}, is given
by the following operator defined on smooth functions on $\bb R_+$
such that $f'(0) = 0$ (this condition corresponds to the Neumann boundary
so that the origin is a reflecting barrier)
\begin{equation}
\label{gen1dim}
Lf(x) =   \frac{1}{2} \sigma^2 x f''(x) + \kappa \, (\theta-x) f'(x)
\end{equation}


A straightforward computation shows that $L$ is symmetric in
$L_2(\bb R_+,d\pi_\nu)$, where $\pi_\nu$ is the Gamma distribution
given in \eqref{misinv}. Note that we use the same notation for
the Gamma distribution and its density.

The generator $L$ can be written in the explicit self-adjoint form as
\begin{equation}
\label{gensa}
L f(x) = \frac{\sigma^2}{2}  \frac{1}{\pi_\nu(x)}
\bigl[ \pi_\nu(x) x f'(x) \bigr]'
\end{equation}
so that the corresponding Dirichlet form is
\begin{equation}
\label{2.5}
D(f) := - \int_0^\infty \! d\pi_\nu (x) f(x) L f(x)
= \frac{\sigma^2}{2} \int_0^\infty \! d\pi_\nu (x) \, x \, f'(x)^2
\end{equation}
We also define the qudaratic form $D_1$ by
\begin{equation}
D_1 (f) = D(f) + \int_0^\infty\! d\pi_\nu (x) \, f(x)^2
\end{equation}
By standard theory, see e.g.\ \cite{Fukushima}, the form defined
by \eqref{2.5} is closable and the associated Hunt process is the solution
to \eqref{CIR1}.  We shall denote by $D_1$ also
the closure of the form defined above and let $\mc D_1$ be its domain.

We now recall that the capacity of an open set $\mc O \subset \bb R_+$
is defined as
\begin{equation}
\label{2.6}
\capa( \mc O) := \inf_{f \in \mc F_{\mc O}} D_1(f)\,,
\quad \quad
\mc F_ \mc O := \big\{ f\in\mc D_1 \, :  \; f(x) \ge 1\,,\;  x\in \mc O \big\}
\end{equation}
For an arbitrary set $B \subset \bb R_+$ the capacity of $B$ is finally defined as
\begin{equation}
\label{2.7}
\capa(B) := \inf_{\mc O \textrm{ open} \: :  \; \mc O \supset B}
\capa(\mc O)
\end{equation}
A classical result, see e.g.\cite[4.3]{Fukushima}
of the potential theory for diffusion processes is that set with
null capacity are never reached; such sets are called \emph{polar}.

\begin{proposition}[Unidimensional Feller condition]
For the Dirichlet form \eqref{2.5},
the origin, i.e.\ the set $\{0\}$, is polar if and only if $\nu \ge 1$.
\end{proposition}

\begin{proof}.
It is convenient to introduce the quadratic form $D_c$, with $c>0$
\begin{equation}
D_c (f) =  D(f) + c ~ \int_0^\infty \! d\pi(x) \, f(x)^2
\end{equation}
and let $\capa_c$ be the associated capacity. Of course a set is polar
with respect to $\capa_c$ if and only if is polar with respect to
$\capa\equiv\capa_1$.
We shall compute the capacity of $[0,\varepsilon)$ for a convenient
choice of $c$.

\p\n
The minimizer for the variational problem defining $\capa_c([0,\varepsilon))$ solves the equation
\begin{equation}
\label{unidim}
\left\{
\begin{array}{ll}
L f(x) - c f(x) = 0 & x \in (\varepsilon,\infty) \\
f(x) = 1 & x \in [0,\varepsilon] \\
f \in L_2([0,\infty),d\pi_{\nu})
\end{array}
\right.
\end{equation}
The differential equation in \eqref{unidim} reads
\begin{equation}
\frac{1}{2} \, \sigma^2 \, x \, f''(x)
+ \kappa (\, \theta-x) \, f'(x) - c \, f(x) = 0
\end{equation}
i.e.\
\begin{equation}
x \, f''(x) + \nu \Big( 1- \frac{x}{\theta} \Big) f'(x)
- \frac{2c}{\sigma^2} \, f(x) = 0
\end{equation}
that, modulo a change of scale, is a confluent hypergeometric differential
equation \cite{Abramowitz}. Instead of using confluent hypergeometric
functions, it is simpler to perform the change of variable
$f(x)=g(x)/\pi_\nu(x)$.
A straightforward calculation gives
\begin{equation}
\label{unidimg}
x \, g''(x) + \Big[ 2-\nu + \nu \frac{x}{\theta} \Big] g'(x)
+  \Big[ \frac{\nu}{\theta} - \frac{2c}{\sigma^2} \Big] g(x) = 0
\end{equation}
We now take advantage of the arbitrariness of $c$ by choosing
$c=( \nu \sigma^2) / (2 \theta)=\kappa$; in this way the solution
of \eqref{unidimg} satisfying the appropriate boundary conditions is
simply given by
\begin{equation}
\label{2.17}
g(x) = \pi_\nu(\varepsilon) \frac{G(x)}{G(\varepsilon)}
\quad\text{ where }\quad
G(x) := \displaystyle \int_x^{\infty} \!dy \, \frac{\pi_\nu(y)}{y}
\end{equation}
Hence
\begin{equation}
\begin{array}{lcl}
  \capa_{\kappa}([0,\varepsilon)]
  &=& \kappa \int_{0}^{\varepsilon}\!dx \,\pi_\nu(x)
  + \frac{\sigma^2}{2} \int_{\varepsilon}^{\infty} \!dx \, \pi_\nu(x)
  \Big\{ x \Big[ \Big( \frac{g(x)}{\pi_\nu(x)} \Big)' \Big]^2  +
  \frac{\nu}{\theta} \frac{g(x)^2}{\pi_\nu(x)^2} \Bigr\}
  \\
  \\
  &=& \pi_\nu([0,\varepsilon))
  +\frac{\sigma^2}{2}  \frac{\pi_\nu(\varepsilon)^2}{G(\varepsilon)}
  + \frac{\sigma^2}{2} \pi_\nu(\varepsilon)
  \Big[ \nu-1 - \frac{\nu}{\theta} \varepsilon \Big]
\end{array}
\end{equation}
As $\capa_{\kappa}(\{0\})=
\lim_{\varepsilon \downarrow 0}{\capa_{\kappa}([0,\varepsilon))}$
it is now easy to check that the capacity of the origin is null when
$\nu>1$ since all terms vanish as $\varepsilon\downarrow 0$;
for $\nu = 1$ the capacity of the origin is still null since
$G(\varepsilon)$ diverges logarithmically.
Finally for $\nu < 1$, by the asymptotic expansion of
$G(\varepsilon)$, see e.g.\ \cite{Abramowitz},
\begin{equation}
f(x) =\frac{g(x)}{\pi_\nu(x)} \simeq \frac{1}{1-\nu} + \Gamma(\nu-1) \,
  x^{1-\nu} \, \Big( \frac{\nu}{\theta}\Big)^{1-\nu}\,,
\qquad  x\simeq 0
\end{equation}
where we used \eqref{2.17} and $\Gamma(\nu) = (\nu-1) \Gamma(\nu-1)$.
It is now simple to check that $\capa_{\kappa}(\{0\})= \kappa (1-\nu )$.
\end{proof}
\qed

\section{Independent CIR processes}
\label{s:3}

In this section we extend the results of the previous one to
the case of independent CIR processes.
Let $\big(\Omega, \mc F, \mc F_t,\bb P\big)$
be a filtered probability space equipped with a standard
$n$-dimensional Wiener process $w=(w^1,\cdots,w^n)$ and
consider the uncoupled system of Ito
equations
\begin{equation}
\label{2.2}
\begin{array}{ccl}
d X^i_t &=& \kappa_i [\theta_i - X^i_t] dt + \sigma_i\, \sqrt{X^i_t} \, dw^i_t
\\
X^i_0&=&x^i_0
\end{array}
\quad \quad  i=1,\dots, n
\end{equation}
As in the one dimensional case we restrict to the case
$\kappa_i,\theta_i,\sigma_i,x^i_0 >0$ ($i=1,\dots,n$) and
set $\nu_i := 2 \kappa_i \theta_i/ \sigma_i^2$.
Of course $X^i_t \ge 0$  for any $t\in\bb R_+$ and any $i=1,\cdots,n$.

The generator of the $n$-dimensional process $X=(X_1,\dots,X_n)$ is
given by the following operator defined for any smooth
functions on $\bb R_+^n$ such that $\partial_{x_i} f(x) = 0$ if $x_i=0$,


\begin{equation}
\label{2.3}
Lf(x) =
\sum_{i=1}^n
\Bigl[ \frac{1}{2}
 \sigma_i^2 x_i \partial_{x_i x_i} f(x) +
\kappa_i (\theta_i - x_i) \partial_{x_i} f(x) \Bigr]
\end{equation}

Since the processes in
\eqref{2.2} are independent it follows  that $L$ is symmetric in
$L_2(\bb{R}_+^n,d\pi_{\ul{\nu}})$, where $\ul{\nu}:= (\nu_1,\ldots,\nu_n)$ and
$d\pi_{\ul{\nu}}$ is the product of $n$ Gamma distributions with parameters
$\nu_i:= 2\kappa_i\theta_i/\sigma_i^2$.
Its density w.r.t.\ the Lebesgue measure on $\bb
R_+^n$ is $\pi_{\ul{\nu}}(x) ={ \prod_{i=1}^{n}} \pi_{\nu_i}(x_i)$, where
$\pi_{\nu_i}$ is as in \eqref{misinv}.

Similarly to the one dimensional case, we address the question of which
condition the parameters should fulfil so that the $n$-dimensional
process $X= (X^1,\cdots,X^n)$ does not hit the origin, {\it i.e.}
when $\sum_{i=1}^n X^i$ does not hit zero.
From the one-dimensional result it follows immediately
that $X^i>0, i=1,\dots,n $ , {\it i.e.} the
processes $X$ does not hit the coordinate axes, iff $\nu_i\ge 1, i=1,\dots,n $.
However a less stringent condition is sufficient to ensure that $X$ does not hit the origin,
namely iff $\sum_{i=1}^n \nu_i \ge 1$.
This result is proven below firstly by a comparison argument and
successively by using capacity theory.


\begin{proposition}[n-dimensional Feller condition]
\label{t:hon}
  The $n$-dimensional process $X := (X^1,\cdots X^n)$ hits the origin
  with positive probability if and only if
$|\ul{\nu}| := \sum_{i=1}^n \nu_i < 1$.
\end{proposition}

\noindent\emph{Proof.}\
We first show that if $\nu = |\ul{\nu}| \ge 1 $
then $\sum_{i=1}^n X_i $ does not hit zero $\bb P$-a.s.
Let $\ol{\kappa} := \max_{i=1,\cdots,n} \kappa_i$ and
introduce $n$ independent processes $Y^i$ as the solution to the equation
\begin{equation}
\label{2.2.1}
\begin{array}{ccl}
d Y^i_t &=& [ \kappa_i\theta_i - \ol{\kappa} Y^i_t] dt
+ \sigma_i\sqrt{Y^i_t} dw^i_t
\\
Y^i_0&=& x^i_0
\end{array}
\quad \quad  i=1,\dots, n
\end{equation}
Since $X^i\ge 0$ and $Y^i\ge 0$, $i=1,\cdots,n$, by a standard result
on one dimensional Ito equations, see e.g.\ \cite[Thm.~IX.3.7]{RevuzYor},
for each $i=1,\cdots,n$ we have $X^i \ge Y^i$ $\bb P$-a.s. It
is therefore enough to prove that $\sum_{i=1}^n Y^i$ does not hit
zero. Let
\begin{equation}
  \label{2.2.2}
  Z_t := \sum_{i=1}^n \frac 1{\sigma_i^2} Y^i_t\,
  \qquad  \qquad  \qquad z:=\sum_{i=1}^n \frac 1{\sigma_i^2} x^i_0
\end{equation}
From Ito's formula we get
\begin{equation}
\begin{array}{lcl}
  Z_t &=& z + \sum_{i=1}^n \frac 1{\sigma_i^2}
  \dint_0^t\!ds\, \big[  \kappa_i\theta_i - \ol{\kappa} Y^i_s \big]
  + \sum_{i=1}^n \frac 1{\sigma_i^2}  \dint_0^t \sigma_i\sqrt{Y^i_s}
  \, dw^i_s \\
  \\
  &=& z + \dint_0^t\!ds\, \Big[ \frac {\nu}2 - \ol{\kappa} Z_s \Big] + M_t
\end{array}
\end{equation}
where $M_t$ is a martingale with quadratic variation
\begin{equation}
  \langle M \rangle_t =
  \sum_{i=1}^n \frac 1{\sigma_i^2} \int_0^t \!ds\, Y^i_s  =
  \int_0^t\!ds\, Z_s
\end{equation}
We thus see that $Z$ solves, in the sense of the associated martingale problem,
the stochastic equation \eqref{CIR1} with $x_0=z$, $\kappa= \ol{\kappa}$,
$\theta= \nu/(2\ol{\kappa})$, and $\sigma=1$. From the result on the one
dimensional CIR process discussed in Section~\ref{sec:2} we then get that if
$ 2\ol{\kappa}\,\nu/(2\ol{\kappa}) = \nu \ge 1$ then the process $Z$
is $\bb P$-a.s.\ strictly positive.

To show that if $\nu<1$ then $X$ hits the origin with positive
probability we argue in a similar way. Let
$\ul{\kappa} := \min_{i=1,\cdots,n} \kappa_i$ and define $\widetilde Y$ as the
solution to \eqref{2.2.1} with $\ol{\kappa}$ replaced by
$\ul{\kappa}$. Then $X^i \le \widetilde Y^i$ a.s., $i=1,\cdots,n$.
Moreover, letting
$\widetilde Z_t := \sum_{i=1}^n \frac 1{\sigma_i^2} \widetilde Y^i_t$, by
the same computation as above, we get that $\widetilde Z $ solves
\eqref{CIR1} with $x_0=z$, $\kappa= \ul{\kappa}$,
$\theta= \nu/(2\ul{\kappa})$, and $\sigma=1$. The result follows.
\qed

\bigskip

\p\n
The Dirichlet form corresponding to the generator $L$ in \eqref{2.3} is given
\begin{equation}
\label{2.5'}
D(f) = - \int \!d\pi_{\ul{\nu}}(x) \,f(x) \, L f(x)
= \dfrac{1}{2} \sum_{i=1}^n  \sigma_i^2
\int \! d\pi_{\ul{\nu}}(x) \: x_i \bigl[ \partial_{x_i} f(x) \bigr]^2
\end{equation}
As in Section~\ref{sec:2}, given $c>0$ we also define the
Euclidean norm $D_c$
\begin{equation}
D_c (f) = D(f) + c \int\! d\pi_{\ul{\nu}}(x) \, f(x)^2
\end{equation}
We shall also denote by $D_c$ the closure of the form defined above
and let $\mc D$ be its domain.
In the next result we prove that if $\sum_{1=1}^n \nu_i \ge 1$ then
the capacity of the origin vanishes. In the next Section we show how
it implies an analogous statement when the CIR processes are not
anymore independent but they are constructed with suitable correlations.

\begin{proposition}
\label{result}
If $\sum_{i=1}^n\nu_i \ge 1$ then the origin  $\{0\}$
is polar for the Dirichlet form \eqref{2.3}.
\end{proposition}

\noindent\emph{Proof.}\
For notation simplicity we consider only the two dimensional case,
$n=2$ and choose $\theta_1=\theta_2=\sigma_1=\sigma_2=1$.
For $\epsilon>0$ set
\begin{equation}
A_\epsilon:=\Big\{ (x_1,x_2)\in\bb R_+^2 \,:\:
\frac{\nu_1}{\nu_1+\nu_2} x_1 + \frac{\nu_2}{\nu_1+\nu_2} x_2 > \epsilon  \Big\}
\end{equation}
we shall construct a function $f_\epsilon:\bb R_+^2 \to \bb R_+$ with
$f_\epsilon=1$ on $\bb R_+^2\setminus A_\epsilon$ such that if $\nu_1+\nu_2\ge
1$ then, for a suitable $c>0$ (hence for all $c>0$) we have
\begin{equation}
\lim_{\epsilon\to 0} D_c(f_\epsilon) = 0
\end{equation}
by the variational definition of the capacity this implies
$\capa_c(\{0\})=0$.

We choose $f_\epsilon(x_1,x_2) = h_\epsilon\big(
\frac{\nu_1}{\nu_1+\nu_2}x_1+\frac{\nu_2}{\nu_1+\nu_2}x_2\big)$ where
$h_\epsilon:\bb R_+ \to \bb R_+$ will be chosen later.
To estimate $D_c(f_\epsilon)$ we perform the linear change of variables
\begin{equation}
\begin{array}{lcl}
r &=& \dfrac{\nu_1}{\nu_1+\nu_2} \, x_1 + \dfrac{\nu_2}{\nu_1+\nu_2} \,x_2 \\
\\
s &=& -\dfrac{\nu_2}{\nu_1+\nu_2} \, x_1 + \dfrac{\nu_1}{\nu_1+\nu_2} \, x_2
\end{array}
\end{equation}
so that
\begin{equation}
\begin{array}{lcl}
x_1=x_1(r,s) &=&
\dfrac{\nu_1+\nu_2}{\nu_1^2+\nu_2^2} \, \nu_1 \, r -
\dfrac{\nu_1+\nu_2}{\nu_1^2+\nu_2^2} \, \nu_2 \, s \\
\\
x_2=x_2(r,s) &=&
\dfrac{\nu_1+\nu_2}{\nu_1^2+\nu_2^2} \, \nu_2 \, r +
\dfrac{\nu_1+\nu_2}{\nu_1^2+\nu_2^2} \, \nu_1 \, s
\end{array}
\end{equation}
We then have
\begin{equation}
\begin{array}{l}
{\displaystyle
D_c(f_\epsilon) = \int\!\!\int_{A_\epsilon}\!dx_1dx_2\: \pi_{\nu_1}(x_1)\pi_{\nu_2}(x_2)
\Big\{
x_1\big[\partial_{x_1} f_\epsilon(x_1,x_2) \big]^2
+ x_2\big[\partial_{x_2} f_\epsilon(x_1,x_2) \big]^2
+ c f_\epsilon(x_1,x_2)^2
\Big\}
}
\\
\qquad
{\displaystyle
= \frac{(\nu_1+\nu_2)^2}{\nu_1^2+\nu_2^2} N_1 N_2
\int_\epsilon^\infty \!dr \: e^{-(\nu_1+\nu_2)r} h_\epsilon'(r)^2
\int_{-\frac{\nu_2}{\nu_1} r}^{\frac{\nu_1}{\nu_2}r}\!ds \:
}
\\
\qquad\qquad\qquad
{\displaystyle
 \Big\{
\frac{\nu_1^2}{(\nu_1+\nu_2)^2}
x_1(r,s)^{\nu_1}x_2(r,s)^{\nu_2-1}
+\frac{\nu_2^2}{(\nu_1+\nu_2)^2}
x_1(r,s)^{\nu_1-1}x_2(r,s)^{\nu_2}
\Big\}
}
\\
\qquad
{\displaystyle
+ c \, \frac{(\nu_1+\nu_2)^2}{\nu_1^2+\nu_2^2} N_1 N_2
\int_\epsilon^\infty \!dr \: e^{-(\nu_1+\nu_2)r} h_\epsilon(r)^2
\int_{-\frac{\nu_2}{\nu_1} r}^{\frac{\nu_1}{\nu_2}r}\!ds \:
x_1(r,s)^{\nu_1-1}x_2(r,s)^{\nu_2-1}
}
\end{array}
\end{equation}
By the change of variable $s=ry$ we have
\begin{equation}
\begin{array}{l}
{\displaystyle
\int_{-\frac{\nu_2}{\nu_1} r}^{\frac{\nu_1}{\nu_2}r}\!ds \:
x_1(r,s)^{\nu_1-1}x_2(r,s)^{\nu_2-1}
}
\\
\qquad
{\displaystyle
=
r^{\nu_1+\nu_2 -1}
\int_{-\frac{\nu_2}{\nu_1}}^{\frac{\nu_1}{\nu_2}}\!dy \:
\Big[ \frac{\nu_1+\nu_2}{\nu_1^2+\nu_2^2} \nu_1
-\frac{\nu_1+\nu_2}{\nu_1^2+\nu_2^2} \nu_2 y \Big]^{1-\nu_1}
\Big[
\frac{\nu_1+\nu_2}{\nu_1^2+\nu_2^2} \nu_2
+\frac{\nu_1+\nu_2}{\nu_1^2+\nu_2^2} \nu_1 y \Big]^{1-\nu_2}
}
\\
\qquad
{\displaystyle
=:r^{\nu_1+\nu_2 -1} C_1
}
\end{array}
\end{equation}
as well as
\begin{equation}
\begin{array}{l}
\dint_{-\frac{\nu_2}{\nu_1} r}^{\frac{\nu_1}{\nu_2}r}\!ds \:
 \Big\{
\frac{\nu_1^2}{(\nu_1+\nu_2)^2}
x_1(r,s)^{\nu_1}x_2(r,s)^{\nu_2-1}
+\frac{\nu_2^2}{(\nu_1+\nu_2)^2}
x_1(r,s)^{\nu_1-1}x_2(r,s)^{\nu_2}
\Big\} \\
\\
=: r^{\nu_1+\nu_2} C_2
\end{array}
\end{equation}
for suitable constants $C_1,C_2>0$ depending only on $\nu_1,\nu_2$. Therefore
\begin{equation}
\begin{array}{l}
{\displaystyle
D_c(f_\epsilon)
}
{\displaystyle
= \frac{(\nu_1+\nu_2)^2}{\nu_1^2+\nu_2^2} N_1 N_2
\int_\epsilon^\infty \!dr \: r^{\nu_1+\nu_2 -1} e^{-(\nu_1+\nu_2)r}
\big\{ C_2 r [h_\epsilon'(r)]^2 + c C_1 h_\epsilon(r)^2 \big\}
}
\end{array}
\end{equation}

\medskip

\p\n
and we conclude the proof by choosing $h_\epsilon$ as in the one
dimensional case with parameter $\nu_1+\nu_2$ for an appropriate $c>0$.
\qed

\section{A class of bidimensional correlated processes}

While multi-dimensional independent CIR processes have been
widely employed to describe systems with mean reverting
characteristics, correlated CIR-like processes are less
popular. A general setting for multi-factor mean-reverting
processes where interest rates and credit spreads are affine
in the state variables has been investigated in the works of
Duffie and Singleton \cite{DuffieSingleton} and
Dai and Singleton \cite{DaiSingleton}, where \emph{e.g.}
the number of state variables is three and
\begin{equation}
dX_t = \kappa (\theta - X_t)dt + \Sigma \sqrt{S(X_t)} \, dW_t
\end{equation}
where $\theta \, \in {\mathbb R}_{+}^3$, and $\kappa$, $\Sigma$ and $S(t)$ are
$3 \times 3$ matrices, out of which the first two are constant while
$S$ is diagonal and affine in the state variables, so that it is
possible to mix Gaussian and CIR-like processes.

Clearly, the choice of the correlation structure depends
on the properties of the system to be modelled.
We introduce here a class of bivariate processes where
the correlation is such that the invariant measure of the joint
process is equal to that of two independent CIR processes.
As a consequence, in the asymptotic state the two processes
decouple. We shall refer to this model as the asymptotically
decoupling correlated (ADC) model.

This approach is different to the introduction of a correlation
on the underlying Wiener processes in \eqref{2.2}. Indeed we
perturb both the martingale part and the drift terms in
\eqref{2.2} in such a way the decoupling holds as
$t \rightarrow \infty$.
On the other hand for finite times the corrections
can still be relevant.
We thus analyze the class of bidimensional processes of the type

\begin{equation}
\label{3.1}
d  \begin{pmatrix} X_1 \\ X_2 \end{pmatrix}
= A(X_1,X_2) dt +
B(X_1,X_2) \,
d \begin{pmatrix} w_1 \\ w_2 \end{pmatrix}
\end{equation}
where $w_1$ and $w_2$ are independent Wiener processes and
we restrict the choice of $A$ and $B$ to second order polynomials in $X_1$ and $X_2$.
Specifically, given the CIR parameters $\kappa_i, \theta_i, \sigma_i >0$, and the correlation
parameters $\epsilon_i \ge 0$ and $\gamma \in [-\sqrt{\epsilon_1\, \epsilon_2},\sqrt{\epsilon_1 \, \epsilon_2}]$
with $i=1,2$, we choose
\begin{equation}
\label{3.2}
\begin{array} {lc}
\text{} \:& A =
\left( \begin{array}{c} A_1 \\ A_2 \end{array} \right) =
\left( \begin{array}{c}
\kappa_1(1+\beta_{1} X_2) [\theta_1 - X_1] +
\kappa_2 \alpha_2 X_1 [\theta_2 - X_2] \\
\kappa_2 (1 + \beta_2 X_1)[\theta_2 - X_2] + \kappa_1 \alpha_1 X_2 [\theta_1 - X_1]
\end{array} \right) \\
\\
\: \text{} \:& B(x_1,x_2) B(x_1,x_2)^T = S(x_1,x_2)
\\
\\
\: \text{with} \:& S(x_1,x_2)= \left(
\begin{array} {cc}
\sigma_1^2 x_1 + \varepsilon_1 x_1 x_2 & \gamma x_1 x_2 \\
\gamma x_1 x_2 & \sigma_2^2 x_2 + \varepsilon_2 x_1 x_2
\end{array} \right) \\
\\
\: \text{and} \:& \beta_i = \dfrac{\varepsilon_i}{\sigma_i^2} ~~~~
\alpha_i = \dfrac{\gamma}{\sigma_i^2} ~~~~ i=1,2
\end{array}
\end{equation}
Thus while $\varepsilon_1$ and $\varepsilon_2$ measure the degree of
correlation in the diagonal terms, $\gamma$ measures the asymmetry
between the two factors. If $\epsilon_1=\epsilon_2=\gamma=0$ the
process reduces to the independent bidimensional CIR process.

\p\n
Since
\begin{equation}
\label{3.3}
\text{det}(S) =
x_1 x_2 (x_1 \varepsilon_2 \sigma_1^2 + x_2 \varepsilon_1 \sigma_2^2 + \sigma_1^2 \sigma_2^2)
+ x_1^2 x_2^2 ( \varepsilon_1 \varepsilon_2 - \gamma^2)
\end{equation}
under the model conditions  $\varepsilon_1 \ge 0$, $\varepsilon_1 \ge 0$
and $\varepsilon_1 \varepsilon_2 \ge \gamma^2$ the diffusion matrix is definite positive.
Moreover the eigenvalues $e_{1,2}$ and eigenvectors $\hat{e}_{1,2}$ of $S$ are
\begin{equation}
\begin{array}{lcl}
e_{1,2} &=& \dfrac{1}{2} \Biggl[ (\sigma_1^2 x_1 + \varepsilon_1 x_1 x_2) +
(\sigma_2^2 x_2 + \varepsilon_2 x_1 x_2) \, \pm \\
& & \sqrt{[(\sigma_1^2 x_1 + \varepsilon_1 x_1 x_2) -
(\sigma_2^2 x_2 + \varepsilon_2 x_1 x_2)]^2 + 4 (\gamma x y)^2} \Biggr]
\end{array}
\end{equation}
\begin{equation}
\begin{array}{lcl}
\hat{e}_{1,2} &=&
\Biggl( \dfrac{1}{2\gamma x y} \Bigl[ (\sigma_1^2 x_1 + \varepsilon_1 x_1 x_2) -
(\sigma_2^2 x_2 + \varepsilon_2 x_1 x_2) \, \mp \\
&&  \sqrt{
[(\sigma_1^2 x_1 + \varepsilon_1 x_1 x_2) -
(\sigma_2^2 x_2 + \varepsilon_2 x_1 x_2)]^2 + 4 (\gamma x y)^2} ~ \Bigr] , 1 \Biggr)
\end{array}
\end{equation}
while, out of the possible representations of the matrix $B$, the simplest
is obtained by canonical Cholesky decomposition
\begin{equation}
\label{3.4}
\begin{array}{lc}
& B = \dfrac{1}{\sqrt{\sigma_1^2 x_1 + \varepsilon_1 x_1 x_2}}
\Biggl(\begin{array} {cc} \sigma_1^2 x_1 + \varepsilon_1 x_1 x_2  & \gamma x_1 x_2 \\
0 & \sqrt{det(S)}
\end{array} \Biggr) \\
\end{array}
\end{equation}

%


The model as defined above shows two relevant features: {\it (a)} the two random variables
$X_1(t)$ and $X_2(t)$ decouple as $t \rightarrow \infty$; and {\it (b)} the processes
$(X_1,X_2)$ does not hit the origin if $\nu_1 + \nu_2 \ge 1$, where as for the uncorrelated
process $\nu_i = 2 \kappa_i \theta_i / \sigma_i^2  (i=1,2)$.

\medskip

We now show that the process $(X_1,X_2)$ is reversible with
respect to the probability measure $\pi_{\nu_1, \,\nu_2}$, which, as previously, is
the product of two independent Gamma distributions of parameters $\nu_1$ and $\nu_2$.
In other words the generator $L$ of $(X_1,X_2)$ can be written as
\begin{equation}
\label{4.7}
L f(x) = \frac{1}{2} \frac{1}{\pi(x)} \nabla \cdot \bigl( \pi(x) S(x) \nabla f(x)  \bigr)
\end{equation}
where  $\nabla$ denotes the gradient and $\nabla \cdot$  the divergence. The proof
\eqref{4.7} amounts to a straightforward computation. The explicit form of the left
hand term in \eqref{4.7} is
\begin{equation}
L = \bigl[ A_1 \partial_{x_1} + A_2 \partial_{x_2} \bigr] +
\frac{1}{2} \bigl[ S_{11} \partial_{x_1 x_1} + 2 S_{12} \partial_{x_1 x_2} + S_{22} \partial_{x_2 x_2} \bigr]
\end{equation}
while the right-hand term is
\begin{equation}
\begin{array}{l}
\dfrac{1}{2} \dfrac{1}{\pi(x)} \nabla \cdot \biggl( \pi(x) S(x) \nabla f(x)  \biggr) =
\\
\\
\begin{array}{lcl}
&=& \dfrac{1}{2} \dfrac{1}{\pi(x)}
\Biggl\{ \partial_{x_1} \Bigl[ \pi(x) (S_{11} \partial_{x_1} + S_{12} \partial_{x_2})f(x) \Bigr] +
\partial_{x_2} \Bigl[ \pi(x) (S_{21} \partial_{x_1} + S_{22} \partial_{x_2}) f(x) \Bigr] \Biggr\}  \\
\\
&=& \dfrac{1}{2} \bigl[ S_{11} \partial_{x_1} \ln{\pi(x)} + \partial_{x_1} S_{11} + S_{21} \partial_{x_2} \ln{\pi(x) + \partial_{x_2} S_{21}}\bigr] \: \partial_{x_1} f(x) \, +
\\
\\
&& \dfrac{1}{2} \bigl[ S_{22} \partial_{x_2} \ln{\pi(x)} + \partial_{x_2} S_{22} + S_{12} \partial_{x_1} \ln{\pi(x)} + \partial_{x_1} S_{12} \bigr] \: \partial_{x_2} f(x) \, +
\\
\\
&& \dfrac{1}{2} \bigl[ S_{11} \partial_{x_1 x_1} + 2 S_{12} \partial_{x_1 x_2} + S_{22} \partial_{x_2 x_2} \bigr] f(x)
\end{array}
\end{array}
\end{equation}
In the expression above the first term in squared brackets is equal to $A_1$
\begin{equation}
\begin{array}{l}
\dfrac{1}{2} \biggl[ (\sigma_1^2 x_1 + \varepsilon_1 x_1 x_2) (\dfrac{\nu_1-1}{x_1} -
\dfrac{\nu_1}{\theta_1})
+ (\sigma_1^2+\varepsilon x_2) + (\gamma x_1 x_2) (\dfrac{\nu_2-1}{x_2}-\dfrac{\nu_2}{\theta_2}) +
\gamma x_1 \biggr] = \\
\\
= \kappa_1 (\theta_1-x_1) + \kappa_1 \dfrac{\varepsilon_1}{\sigma_1^2} x_2 (\theta_1-x_1) +
\kappa_2 \dfrac{\gamma}{\sigma_2^2}(\theta_2-x_2) = A_1
\end{array}
\end{equation}
and similarly the second term is $A_2$, which completes the proof.

\medskip


By using tools from potential theory we now show that if $\nu_1+ \nu_2 \ge 1$
then the ADC process does not hit the origin. The Dirichlet form of the two
factor correlated CIR process is
\begin{equation}
  \label{eq:dfcp}
  D_{\mathrm{corr}} (f) = \dfrac{1}{2} \, \int_{\bb R_+^2} \!
  d\pi_{\nu_1,\nu_2}(x) \: \nabla f (x) \cdot S(x) \nabla f(x)
\end{equation}
We then have
\begin{proposition}
\label{capdep}
Let $\nu_1+\nu_2\ge 1$. Then the origin $\{0\}$ is polar for the
Dirichlet form \eqref{eq:dfcp}.
\end{proposition}

\begin{proof}.
Let $S_0$ be the diffusion matrix of two independent CIR processes
with parameters $\nu_1$ and $\nu_2$, namely
\begin{equation}
  \label{eq:Sindip}
   S_0(x_1,x_2)  = \left(
  \begin{array}{cc}
   \sigma_1^2  x_1 & 0 \\
   0& \sigma_2^2 x_2
  \end{array}
  \right)
\end{equation}
Recalling that the diffusion matrix $S(x_1,x_2)$ for the two factors
correlated CIR process has been introduced in \eqref{3.2} and that
$\epsilon_1\epsilon_2 \ge \gamma^2$, a simple computation shows that
for any $x\in\bb R_+^2$ we have $S_0(x) \le S(x)$. This means that for
each $v\in \bb R^2$ we have $v\cdot(S-S_0)v \ge 0$.
This bound translates directly to a comparison of the associated
Dirichlet form, i.e.\
\begin{equation}
  \label{eq:cdf}
  D_{\mathrm{indip}} (f) \le  D_{\mathrm{corr}} (f)
\end{equation}
where $D_{\mathrm{indip}}$ denotes the Dirichlet form of the
two-factor independent CIR processes \eqref{2.5}.
The statement now follows from the variational characterization of the
capacity, see \eqref{2.6}, and Proposition~\ref{result}. \qed
\end{proof}


\section{Application to interest rate modelling}

We have used the ADC model to investigate the interest-rate spread
between the Government debt
of two selected European Union member states, Germany and Italy,
in the hypothesis that the spread reflects the different market opinions
of their respective credit quality.
Data for the German and Italian term structures are deduced from the average bid-ask
prices of zero coupon bonds and strips of coupon bonds quoted on the market on Oct. 31, 2006
(time $t_0$).
The corresponding interest rates have then been interpolated using a
natural cubic spline at thirty equally spaced values of time to maturity $\tau=1,2, \dots 30$
years to build the two term structures of interest rates $\tau \mapsto i_c(t_0,t_0+\tau)$ $(c=D,I)$ and
the term structure of the spread $\tau \mapsto s(t_0,t_0+\tau) = i_I(t_0,t_0+\tau)-i_D(t_0,t_0+\tau)$.

The result of this procedure is reported in Fig. \ref{strutture} where the
German and Italian term structures are shown together with the term structure
of the spread upwardly shifted by 3.65\%. In this way it easier to compare the
dependence on time to maturity of the three curves.
In addition, the  plot also shows the zero coupon swap term structure
$\tau \mapsto i_{zcs}(t_0,t_0+\tau)$  extracted with the standard
bootstrap technique from the values of annual interest rates swaps
(the swap rates used here are those versus the 6 months Euribor, computed using the 30/360
convention).
Noticeably the spread between the zero coupon swap and the German curves is about
22 basis points and is fairly independent from time to maturity. On the contrary,
the spread between the Italian and the German curves increases with time to maturity
at a rate very similar to the German term structure.
Bid-ask spreads on the term structures are not reported on Fig. \ref{strutture} since
they are all smaller than 3 basis points.

The analysis is based on the following main assumptions:
\emph{(a)} there is no credit risk
loading on German bond prices, noticeably rated Aaa by all main credit agencies;
\emph{(b)} the German term structure and
the spread between the Italian and German rates can be described by
two state variables, respectively the \emph{benchmark risk-free} rate $r_t$
and \emph{credit spread} $s_t$ evolving in time as
\begin{equation}
\label{5.1}
dX = d  \begin{pmatrix} r_t \\ s_t \end{pmatrix}
= A(r_t,s_t) dt +
B(r_t,s_t) \,
d \begin{pmatrix} w_1 \\ w_2 \end{pmatrix}
\end{equation}
where $w_1$ and $w_t$ are standard independent Brownian motions and
the matrices $A(r_t,s_t)$ and $B(r_t,s_t)$ are defined within either
\begin{enumerate}
\item a bivariate CIR model where the two factors are independent (model $1$) so that
\begin{equation}
\label{5.2}
\begin{array} {lr}
A(r_t,s_t) = \left( \begin{array}{c}
\kappa_r (\theta_r - r_t) \\
\kappa_s (\theta_s - s_t)
\end{array} \right) &
B(r_t,s_t) B(r_t,s_t)^T = S(r_t,s_t) =
\left(
\begin{array} {cc}
\sigma_r^2 r_t & 0 \\
0 & \sigma_s^2 s_t
\end{array}
\right)
\end{array}
\end{equation}

\item the ADC model introduced in the previous section (model $2$)
for which

\begin{equation}
\label{5.3}
\begin{array} {lc}
\text{} \:& A(r_t,s_t) =
\left( \begin{array}{c}
\kappa_r(1+\beta_{r} s_t) [\theta_r - r_t] +
\kappa_s \alpha_s r_t [\theta_s - s_t] \\
\kappa_s (1 + \beta_s r_t)[\theta_s - s_t] + \kappa_r \alpha_r s_t [\theta_r - r_t]
\end{array} \right) \\
\\
\: \text{} \:& B(r_t,s_t)  B(r_t,s_t) ^T = S(r_t,s_t)  = \left(
\begin{array} {cc}
\sigma_r^2 r_t + \varepsilon_r \, r_t \, s_t & \gamma \, r_t \, s_t \\
\gamma \, r_t \, s_t & \sigma_s^2 s_t + \varepsilon_s \, r_t \, s_t
\end{array} \right)
\\
\\
\: \text{with} \:& \beta_i = \dfrac{\varepsilon_i}{\sigma_i^2} ~~~~
\alpha_i = \dfrac{\gamma}{\sigma_i^2} ~~~~ i=r,s;
\end{array}
\end{equation}

a particular case of model $2$ is that obtained for $\varepsilon_r \!=\! \varepsilon_s \!=\! \gamma \!=\!0$,
when it collapses to model $1$; we should refer to this particular case as the \emph{degenerate} ADC model
and use it for calibration purposes.

\end{enumerate}
Finally, since in this work the analysis has been restricted to a single
calendar date, we assume that (\emph{c}) the equations of the
two models given above are expressed according to the \emph{risk-neutral} probability measure,
so that for the moment being we do not need to further specify the market price of risk.

The formal setting is inspired to the well-known \emph{fractional recovery
of market value} setting of Duffie and Singleton \cite{DuffieSingleton}, that in turn
is inspired to the recovery rules of \emph{over the counter} derivatives
In this setting the prices at time $t_0$, $P_D(X,t_0,T)$ and $P_I(X,t_0,T)$ ,
of a German \emph{risk-free} and an Italian \emph{risky} zero coupon bond paying one euro in
$T$ are obtained by discounting at the risk-free rate $r(t)$ and at
the effective rate $r(t)+s(t)$ (without loss of generality we have absorbed
the fractional recovery rate in the definition of $s(t)$), that is
\begin{equation}
\begin{array}{l}
P_D(X,{t},T) = {\bf E}^{\mathbb Q}\bigl[\displaystyle  e^{-\int_{t}^T r(u) \, du} \, | \, {\mathcal F}_{t}\bigr] \\
\\
P_I(X,{t},T) = {\bf E}^{\mathbb Q}\bigl[\displaystyle e^{ -\int_{t}^T [r(u) + s(u)] \, du} \, | \, {\mathcal F}_{t}\bigr]
\end{array}
\label{integrals}
\end{equation}
The prices can also be obtained by the hedging argument and according to the Feynman-Kac formula, by
solving the partial differential equation
\begin{equation}
\begin{array}{l}
\label{FK}
\begin{cases}
\begin{split}
\partial_t P_c(X,t,T) +
\sum_{i} A_i(X,t) \partial_{x_i} P_c(X,t,T) +
\frac{1}{2} \sum_{i,j} S_{ij} \partial_{x_i x_j} P_c(X,t,T) = \\
 = [r(t) + \id_c s(t) ] \, P_c(X,t,T)
\end{split} \\
P_c(X,T,T)=1
\end{cases} \\
\begin{array}{cccc}
\text{with} & c=D,I & \text{and} & \id_c = \begin{cases} 0 & \text{if} \, c=D \\ 1 & \text{if} \, c=I \end{cases}
\end{array}
\end{array}
\end{equation}

We recall that the one-factor CIR model admits an analytic solution for the price of the unitary
zero coupon bond \cite{CIRb}
\begin{equation}
P(r_{t},{t},T) = {\bf E}^{\mathbb Q} \bigl[ e^{-\int_{t}^T r(u) du} | \, {\mathcal F}_{t} \bigr] =
f({t},T) \, e^{- g({t},T) \, r_{t}}
\label{bond}
\end{equation}
where
\begin{equation}
\begin{array}{cccccccc}
f(t,T) &=& \biggl[ \dfrac{d \, e^{\phi (T-t)}}{\phi \, (e^{d (T-t)}-1) + d} \biggr]^{\nu} & &
g(t,T) &=& \dfrac{e^{d (T-t)}-1}{[\phi \, (e^{d (T-t)}-1) + d]}
\end{array}
\label{defABCir}
\end{equation}
depend on $\nu=2\kappa \theta/\sigma^2$ and on the so-called Brown-Dybvig parameters $d$ and $\phi$:
\begin{equation}
\begin{array}{cc}
d = \displaystyle \sqrt{\kappa^2 + 2 \sigma^2}, &
\phi = \dfrac{1}{2} (d + \kappa)
\end{array}
\end{equation}
By the independence of the two factors, model $1$ admits an analytic solution for the
price of the unitary zero coupon bonds
\begin{equation}
\begin{array}{l}
P_D(X,t,T) = P(r_t,t,T) \\
P_I(X,t,T) = P(r_t,t,T) P(s_t,t,T)
\end{array}
\label{bond2}
\end{equation}
For model $2$ we have chosen to compute the expectation integrals in \eqref{integrals} using the
Euler-Maruyama scheme for the evolution of $X$ and the Simpson quadrature rule
for the (stochastic) discount factor. The time step $h=0.004$ \rm{years} and the number of simulations
$N=5000$ have been chosen by requiring the difference between the numerical result and
the analytic expression to be smaller than few basis points in the case of the
degenerated correlated model.

The two models have then been calibrated to the observed term structures.
The calibration of model $1$ is done in two steps: first
the four parameters of the risk-free curve are determined on the German data,
and then the four parameters describing the evolution of the spread are calibrated
on the Italian curve, having fixed the risk-free ones. For the second model
we have fitted simultaneously the two curves by minimising the sum of the
squared differences between the values of the risk-free rates and the values of
the spreads. The minimization is performed using
the MatLab \cite{MatLab} {\tt fmincon} routine for model $1$, while for model $2$
we have implemented (in {\tt C}) a procedure using the fast adaptive simulated annealing
algorithm  of Ingber \cite{Ingber} in order to speed up the computation by
avoiding the use of time-expensive numerical derivatives.
The results of the fits are reported in Table 1, while the differences between the
fitted curves and the observed ones are reported in Fig. \ref{fit}.
Notice that the values of the correlation parameters $\epsilon_r$, $\epsilon_s$ and $\gamma$ are
different from zero and the $\nu_{r,s}$ parameters are both greater than one.

%

\begin{table}[!t]
\caption{Results of the fit for model $1$ (bivariate CIR) and model $2$ (ADC model);
for calibration purposes the parameters of the degenerated ADC model (second column) are fixed
to those of model $1$. In the second part of the table the values of
$\nu_{r,s}=2 \kappa_{r,s} \, \theta_{r,s}/\sigma_{r,s}^2$ and
${\omega}_{r,s}=\nu_{r,s}/\theta_{r,s}$ are reported.
}
\label{Tab1}
\begin{tabular}{c r | c r  r  r }
\hline \noalign{\smallskip}
\multicolumn{2}{c|}{\bf model 1}       & \multicolumn{4}{c}{\bf model 2} \\
\multicolumn{2}{c|}{(bivariate CIR)} &  & degenerated & \multicolumn{2}{c}{non-degenerated} \\
\noalign{\smallskip}\hline\noalign{\smallskip}
$r_0$      & 3.46\%  &  $r_0$       & 3.46\%  &    3.39\%  &                        \\
$\kappa_r$ & 0.0398  &  $\kappa_r$  & 0.0398  &    0.0636  &                        \\
$\theta_r$ & 5.44\%  &	$\theta_r$  & 5.44\%  &    4.55\%  &                        \\
$\sigma_r$ & 4.55\%  &	$\sigma_r$  & 4.55\%  &    3.87\%  &                        \\
$-$        &         &	$\beta_r$($\epsilon_r$)   & 0 (0)      &    258 (0.3859)   \\
$s_0$      & 0.04\%  &	$s_0$       & 0.04\%  &    0.19\%  &                         \\
$\kappa_s$ & 4.0049  &	$\kappa_s$  & 4.0049  &    3.3345  &                         \\
$\theta_s$ & 0.29\%  &	$\theta_s$  & 0.29\%  &    0.26\%  &                         \\
$\sigma_s$ & 2.58\%  &	$\sigma_s$  & 2.58\%  &    4.23\%  &                         \\
$-$        &         &	$\beta_s$($\epsilon_s$)   & 0 (0)      &    114 (0.2046)    \\
$-$        &         &  $\gamma$    & 0       &    0.2800  &                         \\
\noalign{\smallskip}\hline\noalign{\smallskip}
$\nu_r$    &  2.0857  & $\nu_r$     &  2.0857  &    3.8728  &                         \\
$\nu_s$    & 35.0593  &	$\nu_s$     & 35.0593  &    9.6116  &		 	    \\
$\omega_r$ &  2.608\% &	$\omega_r$  &  2.608\% &    1.174\% &		 	     \\
$\omega_s$ &  0.008\% &	$\omega_s$  &  0.008\% &    0.027\% &		 	     \\
\hline
\end{tabular}
\end{table}

The accuracy in the description of the German term structure is approximately
five basis points, which has to be compared with the maximum bid-ask
spread of about three basis points. On the other hand the description of the Italian
term structure is less accurate with deviations ranging up to approximately
twenty basis points.
Although the
two fitted parameter sets are  different (for example $\theta_r$ in
model $1$ is $5.44\%$ while in model $2$ is $4.55\%$),
the two models show a very similar degree of accuracy.
Analogously,  the structure of the deviations
shown in Fig. \ref{fit} is very similar, possibly indicating the presence of a
missing extra factor to be included in the models.

Figures 3 and 4 show the comparison between the distributions of $r_t$ and $s_t$ at
$\tau=5$ years and $\tau=30$ years computed with the two models, both when the ADC model is
degenerate (showing the quality of the calibration) and when the ADC model is
non-degenerated. The plots show that in both cases the spread $s_t$ has essentially reached
the asymptotic distribution already at $t=5$ years. On the contrary the risk-free rate
$r_t$ shows a slower convergence, particularly in the case of model $2$.
The similarity of the deviations in Fig. \ref{fit} is presumably due to the
\virgolette{fast} decoupling of the two factors in model $2$.

As a final comment, we notice that although our results show that there is no substantial
gain in the description of the term structures by using model $2$ with respect
to model $1$, the inclusion of the correlations in model $2$ modifies the
asymptotic state. This is better appreciated in Fig. 5 where the joint density at
$\tau=30$ years of the two state variables is shown for model $2$ and in Fig. 6 where
the difference between the joint density of the two models is reported.
Qualitatively, this is in agreement with the results found in analysing cross
sections of a single term structure with an unidimensional CIR model, where it is well known
(see, \emph{e.g.}, \cite{DM}, p. 100)  that different sets of parameters can provide very similar quality
of description.
To analyse to what extent the value of
risk measures for portfolios composed of Italian and German bonds are
affected,
it is then necessary to specify risk premia for the ADC model and
calibrate their values by the analysis of historical time series of bond prices.



{\begin{figure} [!h]
\begin{center}
\mbox{\includegraphics[scale=0.6]{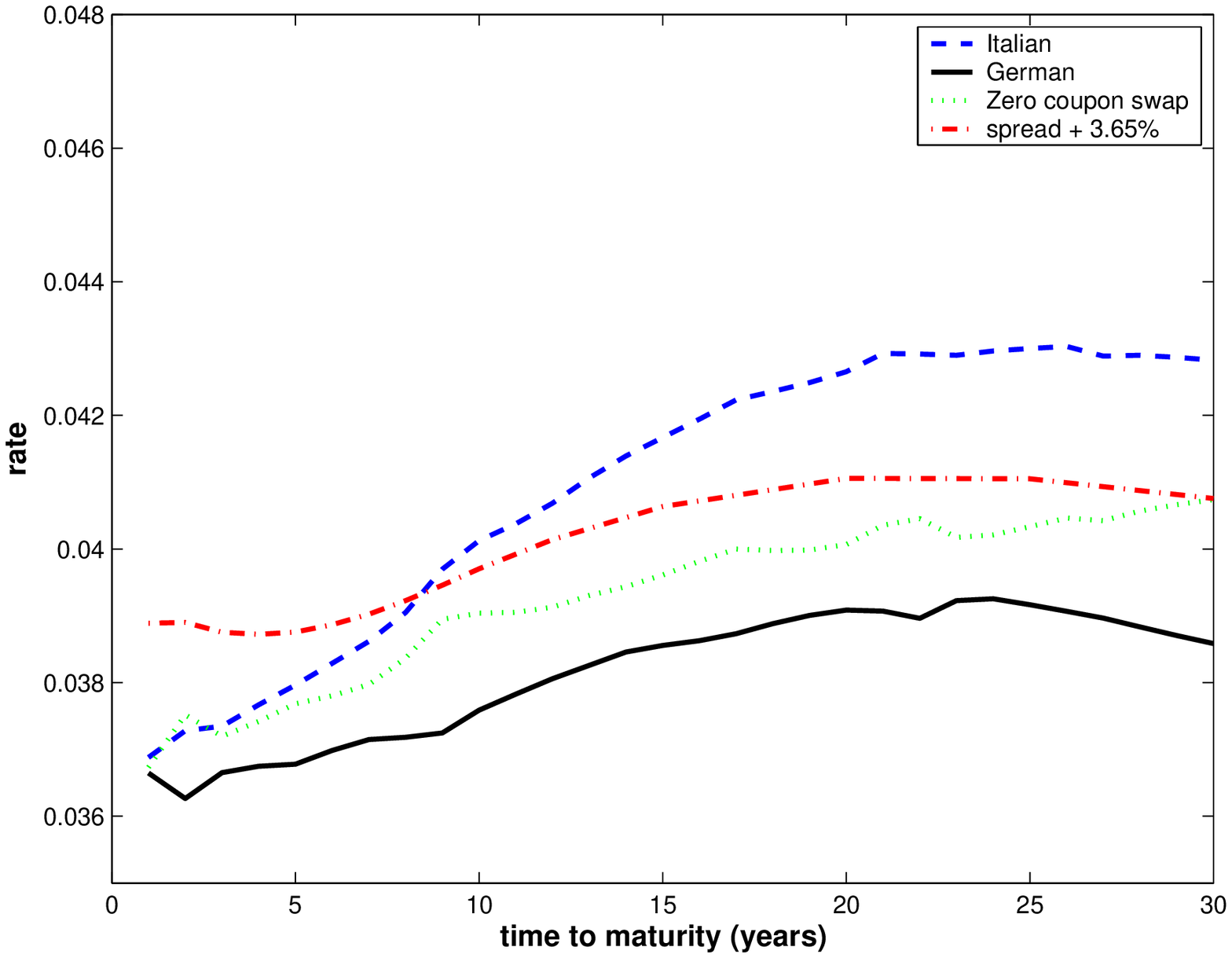}}
\end{center}
\caption{Term structure of the German (continuous line) and Italian (dashed line) interest rates.
The two curves are obtained by interpolating the values derived from the bid-ask average quotations
on Oct. 31, 2006 of zero coupon and strips of government coupon bonds using a natural cubic spline.
The spread (dot-dashed line) between the two curves, shifted by 3.65\% to ease the comparison with the German interest
rate term structure, and the zero coupon swap curve (dotted line) are also plotted.}
\label{strutture}


\begin{center}
\mbox{\includegraphics[scale=0.6]{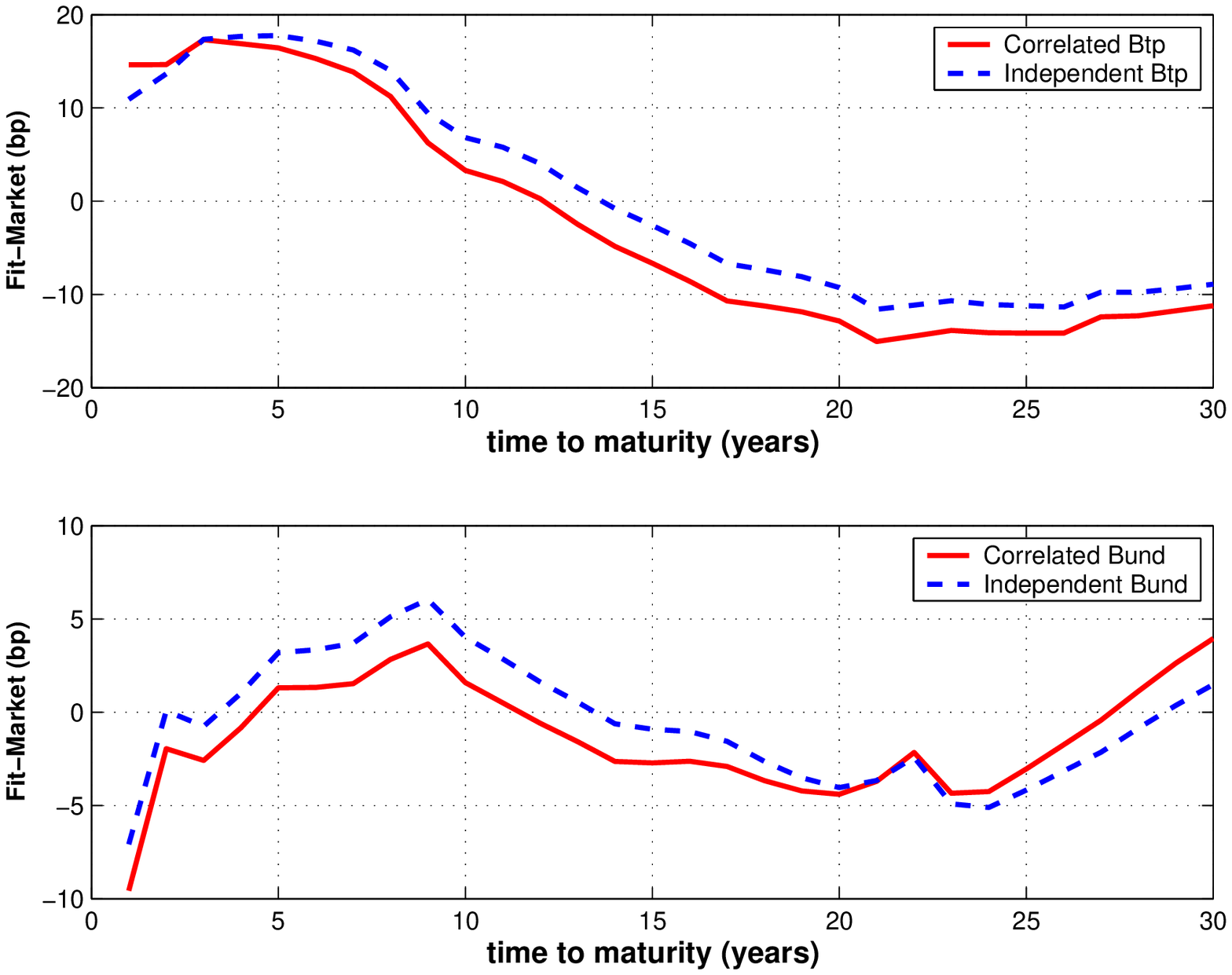}}
\end{center}
\caption{Difference (in basis points) between the fitted term structure and the observed
one for the bivariate model (continuous line) and the ADC model (dashed line) for
the Italian term structure (upper plot) and the German term structure (lower plot).}
\label{fit}
\end{figure}}


\begin{figure*} [!h]
 \label{distrib1}
 \begin{minipage}[b]{5.75cm}
   \centering
   \includegraphics[width=5.75cm]{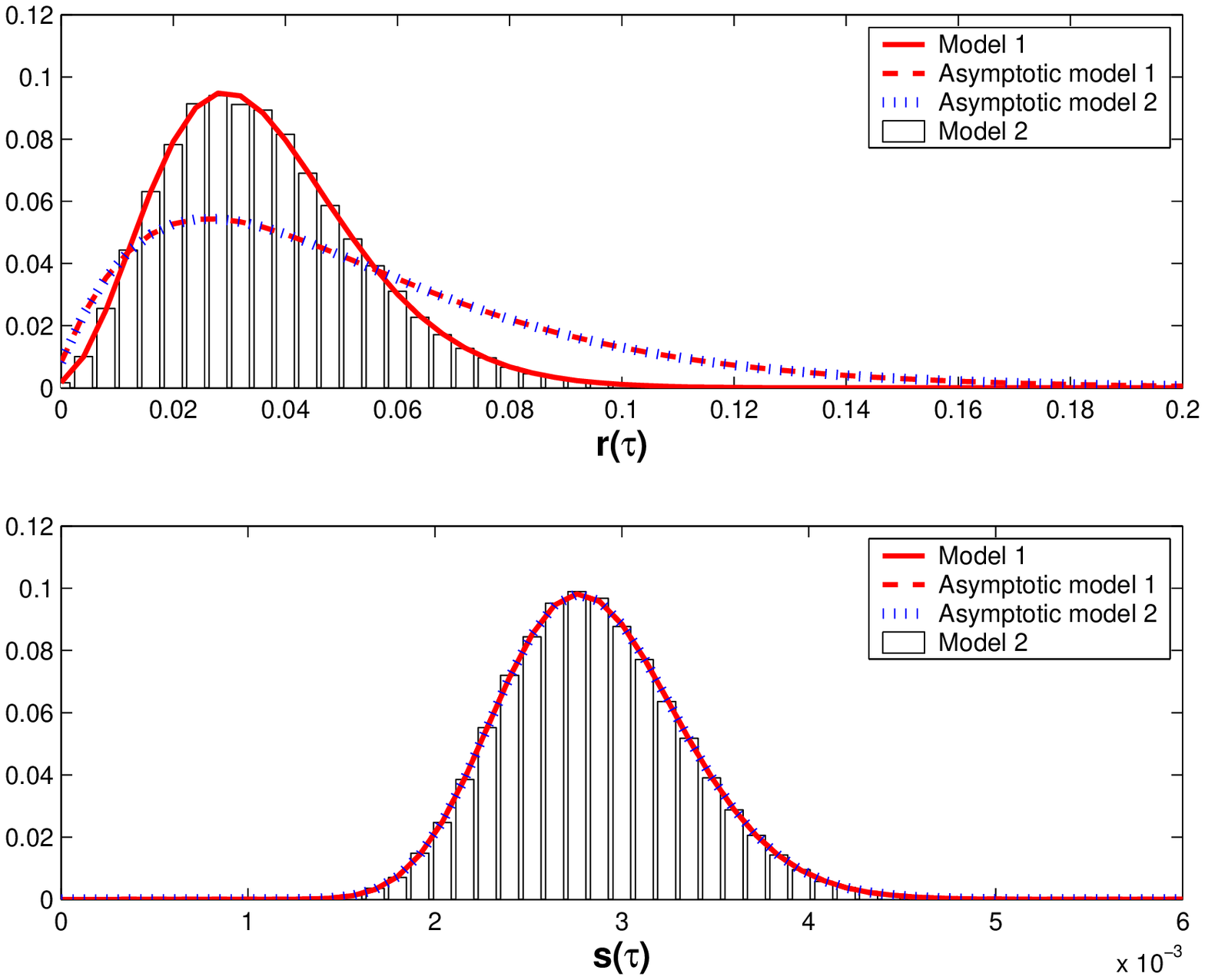}
 \end{minipage}
 \ \hspace{2mm} \hspace{3mm} \
 \begin{minipage}[b]{5.75cm}
  \centering
   \includegraphics[width=5.75cm]{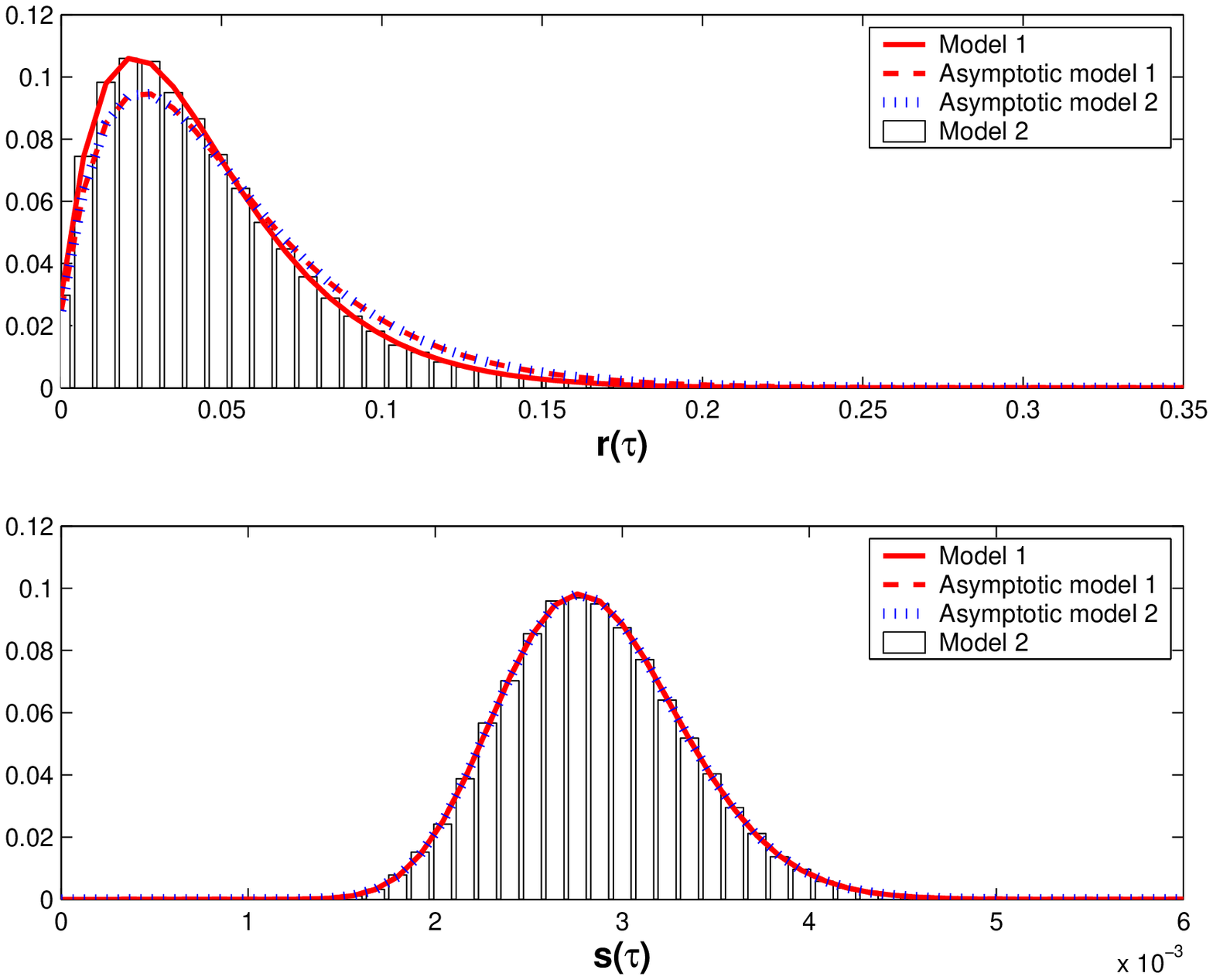}
 \end{minipage}
    \caption{Risk neutral distribution of $r_t$ (upper plots) and $s_t$ (lower plots)
    at $\tau=5$ years (left plots) and $\tau=30$ years (right plots) for the
     model $1$ (continuous line) and the degenerate ADC model (histogram)
    estimated with 100000 Monte Carlo simulations.
    The asymptotic
    value of the two distributions are also drawn (resp. dashed and dotted lines).}
\end{figure*}


\begin{figure*}
 \label{distrib2}
 \begin{minipage}[b]{5.75cm}
   \centering
   \includegraphics[width=5.75cm]{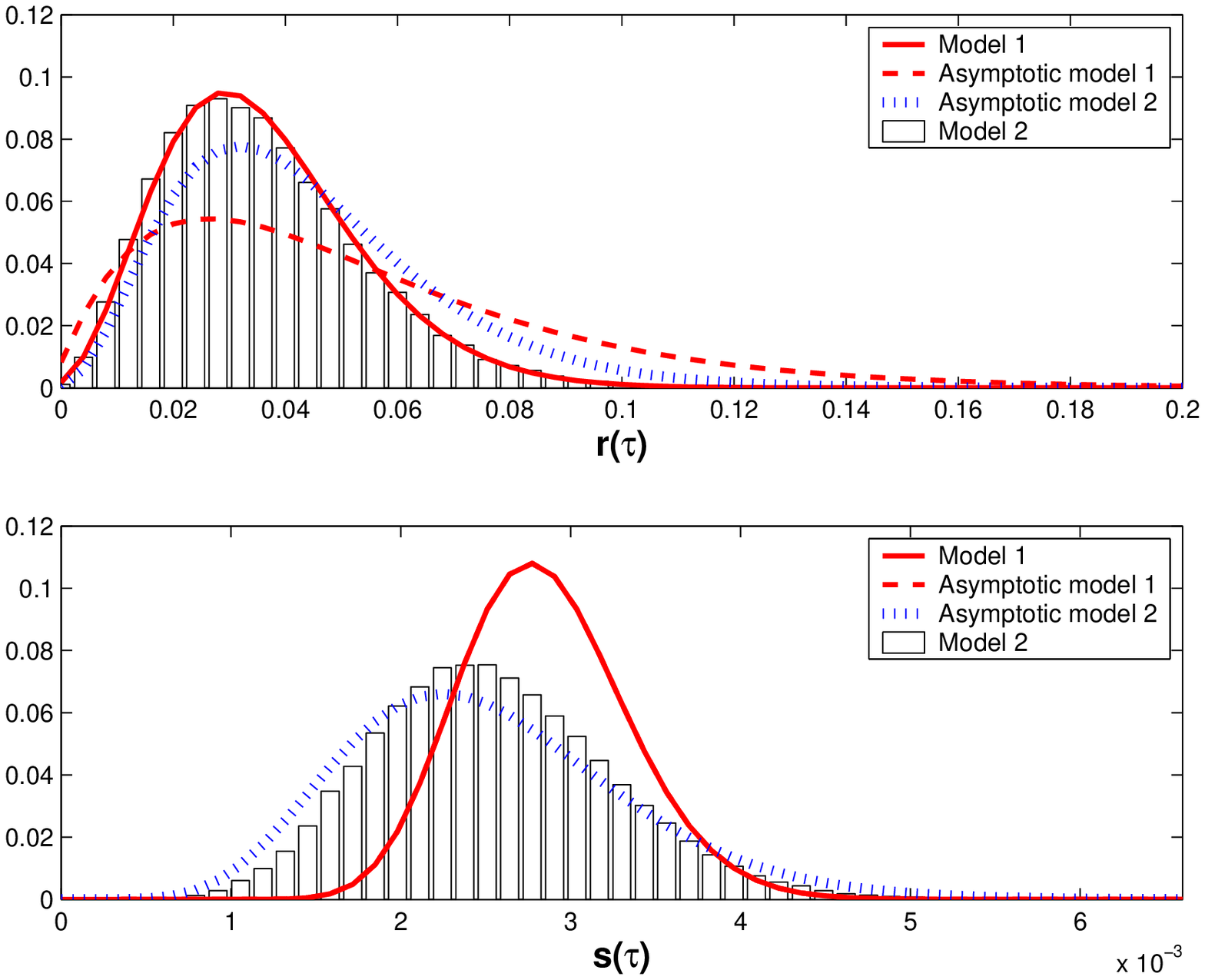}
 \end{minipage}
 \ \hspace{2mm} \hspace{3mm} \
 \begin{minipage}[b]{5.75cm}
  \centering
   \includegraphics[width=5.75cm]{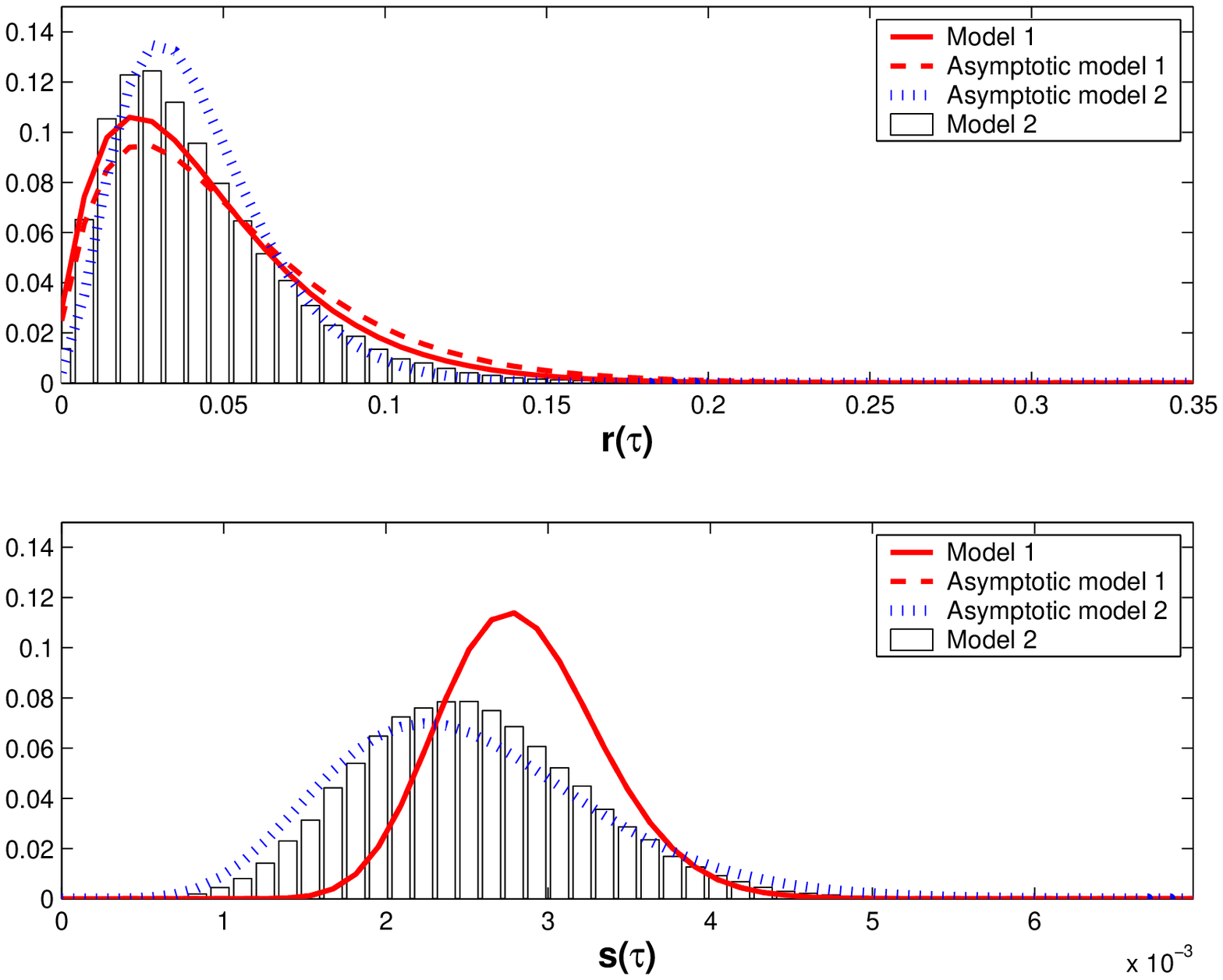}
 \end{minipage}
    \caption{Risk neutral distribution of $r_t$ (upper plots) and $s_t$ (lower plots)
    at $\tau=5$ years (left plots) and $\tau=30$ years (right plots) for the
    model $1$ and the ADC model (histogram)
    estimated with 100000 Monte Carlo simulations. The asymptotic
    value of the two distributions are also drawn (resp. dashed and dotted lines).
    Notice that for both models the distribution of $s_t$ essentially coincides with
    the corresponding asymptotic one already at $t=5$ years.}
\end{figure*}

{\begin{figure} [!h]
\begin{center}
\mbox{\includegraphics[scale=0.6]{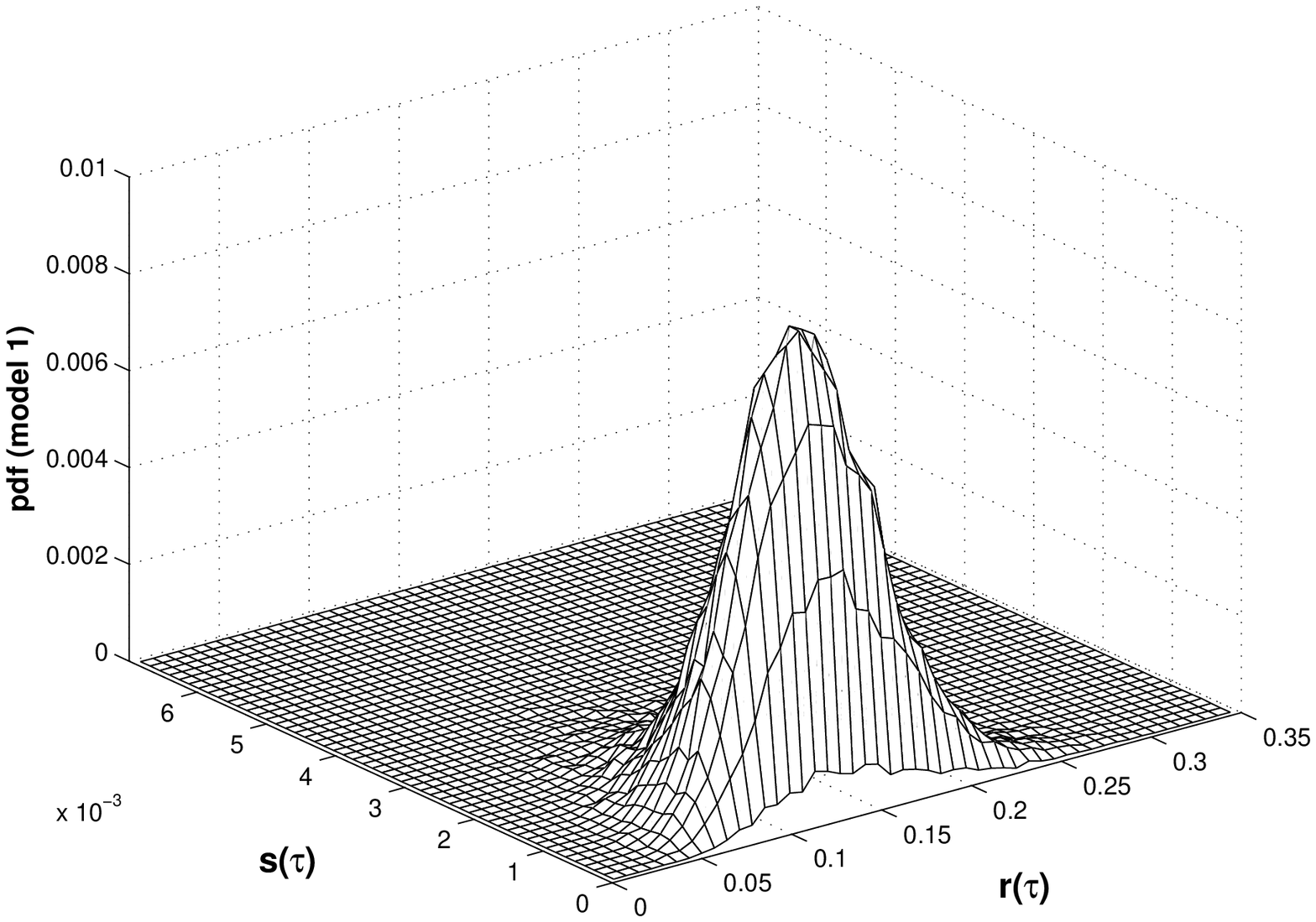}}
\end{center}
\caption{Joint probability density distribution of $(r_t,s_t)$ at $\tau=30$ years for model
$2$ (obtained with 100000 Monte Carlo simulations).}
\label{pdf1}


\begin{center}
\mbox{\includegraphics[scale=0.6]{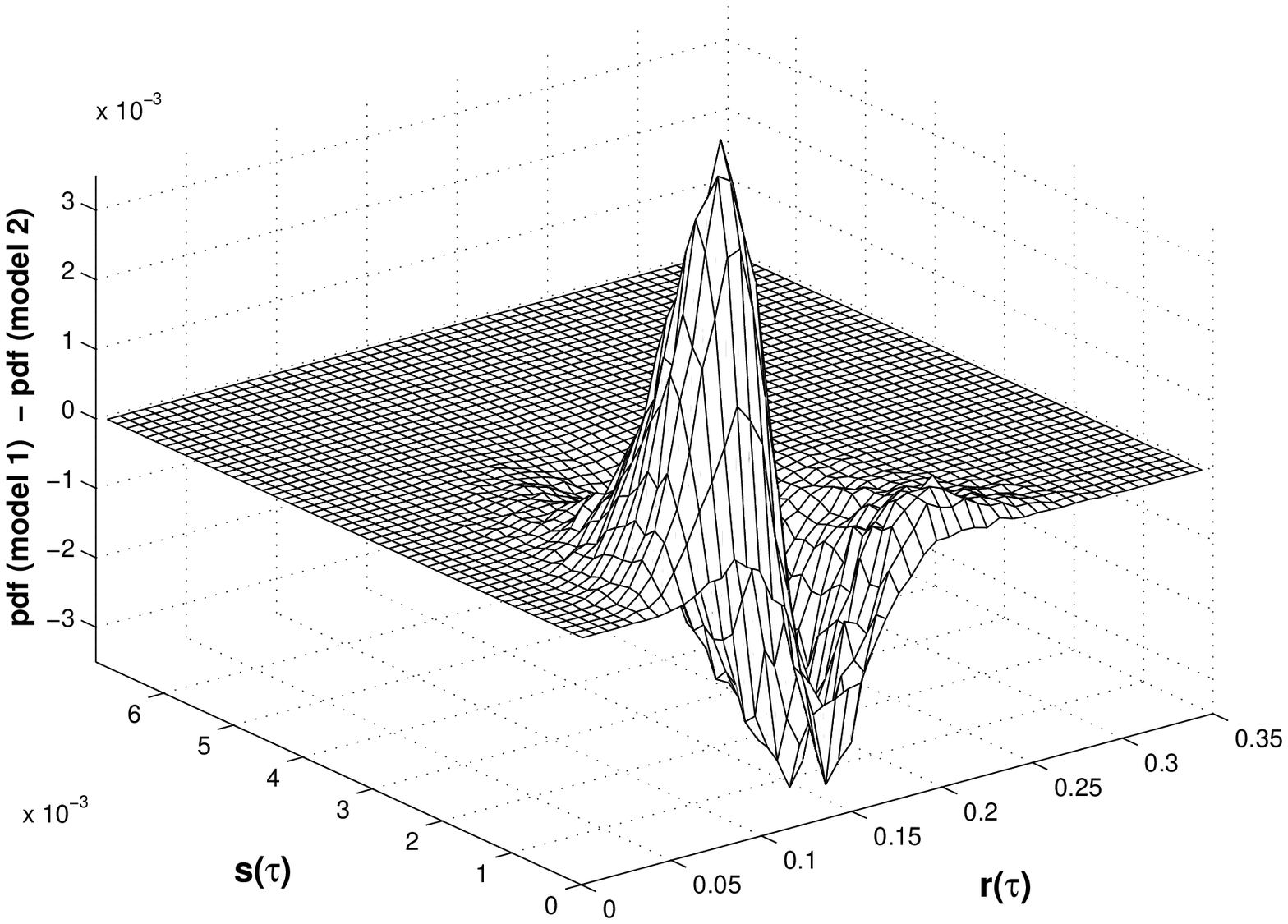}}
\end{center}
\caption{Difference between model $2$ (obtained with 100000 Monte Carlo simulations) and model $1$
in the joint probability density distribution
of $(r_t,s_t)$ at $\tau=30$ years.}
\label{pdfdelta}
\end{figure}}



\vfill
\eject






\end{document}